\documentclass{aa}  
\usepackage{graphicx}
\usepackage{float}
\usepackage{array}
\usepackage[pdftex]{hyperref}
\usepackage{placeins}
\usepackage{amsmath}

\usepackage{txfonts}
\usepackage{xcolor}
\begin{document} 

   \title{Near-infrared spectroscopic characterization of the Pallas family}

   \author{P. Chavan (\raisebox{-0.3ex}{\includegraphics[height=10pt]{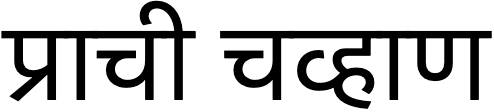}})\inst{1}
        \and B. Yang\inst{1,2}
        \and M.~Bro\v{z}\inst{3}
        \and J.~Hanu{\v s}\inst{3}}

   \institute{Instituto de Estudios Astrofísicos, Facultad de Ingeniería y Ciencias, Universidad Diego Portales, Santiago, Chile\\
        \email{prachi.chavan@mail.udp.cl}\\
        \and
        Planetary Science Institute, 1700 E Fort Lowell Rd STE 106, Tucson, AZ 85719, United States
        \and
        Charles University, Faculty of Mathematics and Physics, Institute of Astronomy, V~Hole{\v s}ovi{\v c}k{\'a}ch 2, 18000 Prague, Czech Republic\\}

 
  \abstract
  {Asteroid families hold clues to the collisional processes that shaped the Solar System over billions of years. The Pallas collisional family, named after (2) Pallas, is notable for its high orbital inclination and the distinct blue color of Pallas and a few larger B-type family members. While Pallas itself, as one of the largest asteroids, has been studied in detail, most of its smaller family members still remain unexplored.}
  {This study aims to characterize the physical properties of medium- to small-sized Pallas family asteroids to investigate the origin of their unusual blueness. We seek to establish connections between asteroid spectra and meteorite analogs. Additionally, we explore the relationship between the Pallas family and the near-Earth object (NEO) (3200) Phaethon.}
  {We conducted near-infrared (NIR) spectroscopy with the NASA Infrared Telescope Facility (IRTF) to collect reflectance spectra for 22 asteroids, including one from the IRTF Legacy Archive. Spectroscopic and dynamical analyses were carried out to identify outliers, while additional data from NEOWISE and \textit{Gaia} were incorporated to examine potential correlations among their physical properties. Meteorite analogs were identified through $\chi^2$ matching using samples from the RELAB database.}
  {The observed Pallas family asteroids exhibit nearly identical spectral profiles, suggesting a homogeneous composition of ejected material. Small variations in spectral slopes are observed, which may result from different levels of alteration experienced by individual asteroids, with some influence from variations in grain size. Most of the observed spectra of the Pallas asteroids, from 0.8 to 2.2 $\mu$m, closely resemble those of the CY and CI meteorites. The blueness of asteroid surfaces is likely due to the presence of magnetite, troilite, or phyllosilicates, which are products of aqueous alteration. The striking spectral similarity between (3200) Phaethon and Pallas family members of comparable sizes suggests a potential common origin.}
  {}


   \maketitle
   
\section{Introduction} \label{sec:intro}

Asteroids, small rocky objects orbiting the Sun, are considered remnants of the formation of our Solar System, dating back over 4.5 billion years \citep{DeMeo:2014}. The violent collisions among these bodies not only shaped the early environment of the inner planets but also played a significant role in delivering water and organic compounds, contributing to the development of life on Earth \citep{Bottke:2015, Osinski:2020}. Following a collisional breakup or cratering event of a large asteroid, the ejected fragments partly reaccumulated and remained as individual asteroids sharing similar orbital elements, in particular, the semimajor axis, the eccentricity, and the inclination. These groups of asteroids are commonly referred to as asteroid families \citep [and refs. therein]{NesvornyIdentificationDynamicsOfAsteroidFamilies,Nesvorny_2024ApJS..274...25N}.

The formation and evolution of asteroid families is a complex process influenced by many factors, including the size, internal structure, and composition of the parent body, as well as the dynamical and physical properties of the impactor \citep{ Michel:2004, Durda:2007,NesvornyIdentificationDynamicsOfAsteroidFamilies}. Studying these families allows us to investigate the physics of impacts at a scale that is impossible to reproduce in laboratories. Such studies provide an understanding of the complex interplay of forces that have shaped our Solar System over billions of years \citep{Bottke:2014, Masiero:2015, Novakovic:2022}. By measuring the physical properties of individual asteroid family members, we can learn about the properties of the respective parent body and, by extension, about the diversity of primordial planetesimals formed throughout the Solar System.

The Pallas family is located at 2.75 au in the main asteroid belt, with an unusually high inclination of 34.8$^{\circ}$ \citep{ViolentHistory(2)Pallas}, which makes it challenging for a spacecraft to reach. Understanding the origin of their highly excited orbits could provide important clues to the processes that have shaped the asteroid belt over billions of years. The family was first noted by Kiyotsugu Hirayama in 1928 \citep{KiyotsuguHirayama} whereas \cite{Lemaitre1994} were the first to confirm the existence of a possible collisional family around (2) Pallas. The family consists of almost 300 members \citep{Broz:2024} and is believed to be formed via a cratering event \citep{SpectroscopyOfB-typeAsteroids}. This event produced ejecta of fragments smaller than 20 km while preserving the bulk volume of Pallas \citep{ViolentHistory(2)Pallas}.

As one of the largest asteroids, (2) Pallas has been a target of interest for decades.  It is identified as a primitive asteroid within the C-complex \citep{Bus1999, DeMeo:2009}, which is hypothesized to have formed farther from the Sun, experiencing lower temperatures and fewer alteration processes. As a result, these asteroids are considered more pristine than other classes, potentially preserving key information about the early Solar System's formation and evolution. Pallas itself and some of its family members are classified as B-type asteroids, which are notably rare, accounting for only 4\% of the known asteroid population \citep{BinMagnetite} and showing a distinctive blue color in the optical relative to the Sun. To date, no satisfactory explanation has been found for this unique blueness. In the infrared, some B-type asteroids, including (2) Pallas, show distinctive 3-$\,\mu{\rm m}$ absorption features \citep{Takir2012, Rivkin:2015}, which have been suggested to be associated with hydrated minerals \citep{Hamilton2019}. \cite{2016AliLagoaPCFBtype} find that Pallas family members, on average, tend to exhibit significantly higher visible geometric albedos (12-17\%) than other B-types ($7 \pm 3\%$). This notable difference raises intriguing questions about the composition and surface properties of these high-albedo asteroids, prompting further investigation to understand the origin of their bright nature.

A key approach to addressing such questions involves linking asteroid classes with meteorite classes \citep[e.g.,][]{Eschrig2021, Kramer2021, ConnectingAstMeteo2022DeMeo, Marsset_2024Natur.634..561M, Broz:2024}. These efforts provide a critical framework for connecting the compositional information of meteorites to their parent asteroids, offering a valuable understanding of their physical and chemical properties. (2) Pallas has been associated with different types of meteorites depending on the wavelength range of the spectra. For example, it exhibits the 3-$\mu$m hydration feature similar to CM chondrites, although they have distinct spectral profiles in the visible and near-infrared wavelengths (0.4–2.5 $\mu$m) \citep[and references therein]{ViolentHistory(2)Pallas}. (2) Pallas does not show the 0.7–0.9 $\mu$m absorption features as commonly seen in the CM chondrites, and also, it is bluer and brighter than most of the CM chondrites. In contrast, studies by \cite{SpectroscopyOfB-typeAsteroids} suggest that CK meteorites tend to be the best fit for (2) Pallas, although the 1-$\mu$m absorption band exhibited by these meteorites is absent in Pallas. In addition to that, the 3-$\mu$m hydration feature is absent in the phyllosilicate‐free CK meteorites. In contrast to extensive studies conducted on (2) Pallas itself, only a handful of asteroids from the Pallas family have been studied in the near-infrared (NIR) \citep{SpectroscopyOfB-typeAsteroids, NIRbtypeast}, and most of the members are still unexplored. Therefore, analyzing asteroid-meteorite connections using the NIR spectra of more members of the Pallas family would significantly improve our understanding of (2) Pallas and Pallas-like asteroids.

In addition to the Pallas family, there is growing interest in studying B-types and their connections to the near-Earth population. This interest is largely driven by recent sample-return missions to (162173) Ryugu and (101955) Bennu, as well as planned future missions to (3200) Phaethon \citep{OSIRISmission, Watanabe_2019Sci...364..268W, RyuguSampleIvuna2023, RyuguSampleFormationEvolution2023}. Phaethon is a notable near-Earth object (NEO) with a very low perihelion distance of 0.14\,au and is a source of the Geminid meteoroid stream \citep{Whipple:1983, Williams:1993}. It is classified as a B-type object due to its negative reflectance slope \citep{DeMeo:2009}. It has long been noted that the spectrum of Phaethon resembles those of (2) Pallas and some members of the Pallas family in the visible and near-infrared wavelengths \citep{Jewitt:2006, OriginofPhaethon}. \cite{OriginofPhaethon} proposed a dynamical link between Phaethon and the Pallas family, suggesting that asteroids escaping from this family could populate the highly inclined regions of near-Earth space. However, recent studies argued that the orbit of Phaethon, with its high eccentricity ($e = 0.89$), more likely originates from the Polana family \citep{Broz:2024}.

In this study, we present the largest sample of 23 Pallas family asteroids, including (2) Pallas, observed from 0.8 to 2.5\,$\mu$m, aiming to characterize their physical properties and better understand the distinctive blue spectral slope of B-type asteroids. This paper is organized as follows. In Section~2, we describe the observational data and reduction techniques. In Section~3, we analyze the spectral properties of the Pallas family and search for meteorite analogs to constrain their composition and surface properties. In Section~4, we discuss potential causes of the negative spectral slope in B-type asteroids and propose plausible hypotheses for the spectral features of the Pallas family. In Section~4.1, we revisit the origin hypothesis for (3200) Phaethon, highlighting its striking similarity to two Pallas family asteroids of comparable size. Suggestions for future observations are presented at the end of Section~4. Finally, a summary of the major findings is presented in Section~5.

\section{Observations and data reduction}

Near-infrared (NIR) spectroscopy was carried out at the 3.2 m NASA Infrared Telescope Facility (IRTF) atop Mauna Kea in Hawaii over six observing runs between 2010 and 2012. For each run, a target list was selected from \cite{Nesvorny:2010}, based on the brightness and observability of Pallas family members. Observations were taken using the SpeX instrument \citep{SpeXinstrument}, a medium-resolution spectrograph that covers a wavelength range of 0.7 to 5.3 $\mu$m. Observations were taken in the prism mode covering a wavelength range of 0.7--2.5 $\mu$m, with the 0.8$\times$15" slit aligned along the parallactic angle for all targets. At least one G-type standard star with a similar airmass was observed for each asteroid to correct for both the solar spectrum and telluric absorption. When multiple stars were available, the star with the smallest airmass difference or the least residual telluric correction was used to compute the relative reflectance. An IDL-based spectral reduction program, SpeXtool \citep{Spextool}, was adopted to perform standard data reduction steps like flat fielding, wavelength calibration, spectrum extraction, and spectra combination. An atmospheric model, ATRAN \citep{ATRAN}, was applied to fit and remove the strong absorption features in the NIR spectra due to the Earth's atmosphere. Table~\ref{tab:1} lists the observation dates, basic observational parameters, airmass of the objects, and the corresponding G-type standard stars. Furthermore, we searched the IRTF Legacy Archive for observations of Pallas family members, based on \cite{Nesvorny:2015}, and included the spectrum of (5690) 1992 EU.

    \begin{table*}
    \caption{\label{tab:1}Observational parameters of 22 objects.}
    \centering
        \begin{tabular}{ccccccccc}
        \hline\hline
        Asteroid & UT Date & $r_h$ & $\Delta$ & $\Phi$ & V & Airmass & Standard star\\
        - & - & (AU) & (AU) & (deg) & (mag) & - & -\\
        \hline
        3579 & 2010--Mar--29 & 2.132 & 1.144 & 5.11 & 16.22 &1.19 & SA105-56\\
        4969 & 2009--Aug--08 & 2.324 & 1.610 & 21.44 & 16.83 & 1.02 & HD218633\\
        5222 & 2010--Mar--28 & 2.373 & 1.470 & 13.05 & 15.07 & 1.89 & HD10086\\
        5234 & 2010--Mar--29 & 3.177 & 2.275 & 9.11 & 17.18 & 1.52 & HD114964\\
        5690* & 2009--Oct--18 & 2.217 & 1.325 & 14.86 & 16.38 & 1.06 & HD223238\\
        25853 & 2012--Nov--05 & 2.267 & 1.401 & 15.57 & 16.93 & 1.34 & HD19328\\
        33750 & 2010--Mar--29 & 1.989 & 1.104 & 17.93 & 15.55 & 1.39 & HD106003\\
        46037 & 2011--Sep--27 & 1.981 & 1.193 & 23.27 & 16.98 & 1.15 & HD180883\\
        57050 & 2012--Jan--03 & 1.777 & 1.285 & 32.72 & 17.00 & 1.30 & HD15627\\
        58296 & 2012--Jul--22 & 2.061 & 1.447 & 27.08 & 17.85 & 1.03 & HD182081\\
        66714 & 2010--Mar--28 & 2.261 & 1.291 & 7.83 & 17.19 & 1.29 & SA105-56\\
        66803 & 2012--Jul--22 & 3.124 & 2.313 & 13.05 & 18.05 & 1.42 & HD154805\\
        66846 & 2012--Jul--19 & 1.926 & 1.186 & 26.63 & 17.07 & 1.13 & BD+303643\\
        68299 & 2012--Nov--06 & 2.162 & 1.275 & 15.30 & 16.97 & 1.15 & HD9729\\
        69371 & 2012--Nov--05 & 2.265 & 1.494 & 19.51 & 17.40 & 1.16 & HD224383\\
        87306 & 2011--Jan--05 & 3.188 & 2.215 & 3.16 & 18.00 & 1.01 & HD42807\\
        90368 & 2013--Sep--02 & 1.835 & 1.153 & 29.61 & 16.73 & 1.24 & HD212809\\
        101283 & 2012--Nov--06 & 2.617 & 1.675 & 8.48 & 18.02 & 1.03 & HD12846\\
        109640 & 2010--Aug--12 & 1.899 & 1.327 & 30.53 & 17.32 & 1.16 & BD+45 411\\
        112876 & 2011--Sep--27 & 2.285 & 1.375 & 13.62 & 18.29 & 1.19 & HD197759\\
        136038 & 2011--Sep--27 & 2.038 & 1.116 & 15.06 & 17.63 & 1.25 & HD197759\\
        153652 & 2011--Jan--05 & 2.202 & 1.287 & 12.32 & 16.77 & 1.08 & BD+41 309\\
        \hline
        \end{tabular}
        \tablefoot{$r_h$ and $\Delta$ are the heliocentric and geocentric distances, respectively, $\Phi$ is the solar phase angle, and V is the estimated visual magnitude. *The spectrum of (5690) 1992 EU is taken from IRTF Legacy Archive dated 2009--10--18 and observed by Thomas, C.}
    \end{table*}

\section{Results}
\subsection{Spectral profiles}

\begin{figure*}
    \centering
    \includegraphics[width=8.95cm]{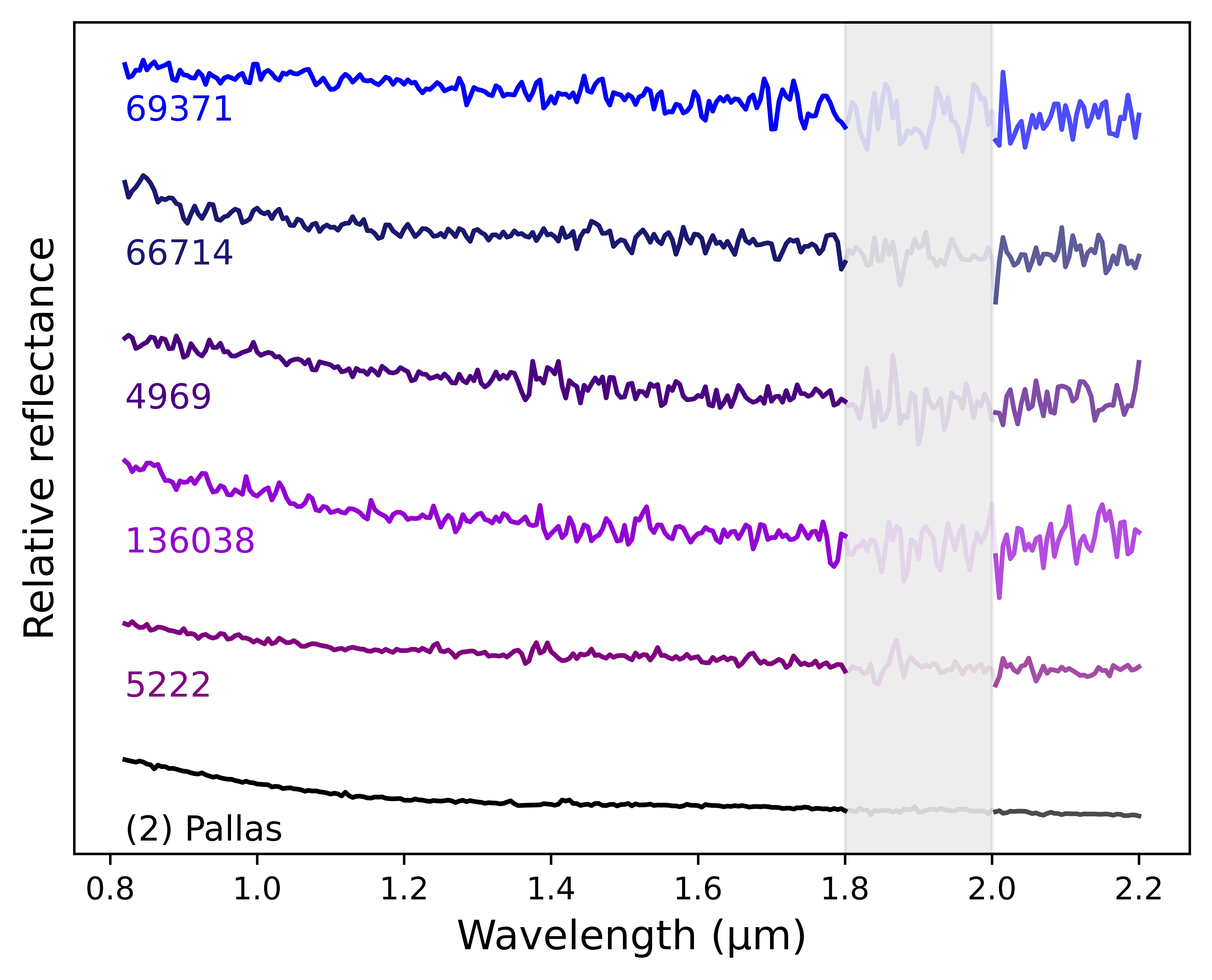}
    \includegraphics[width=8.95cm]{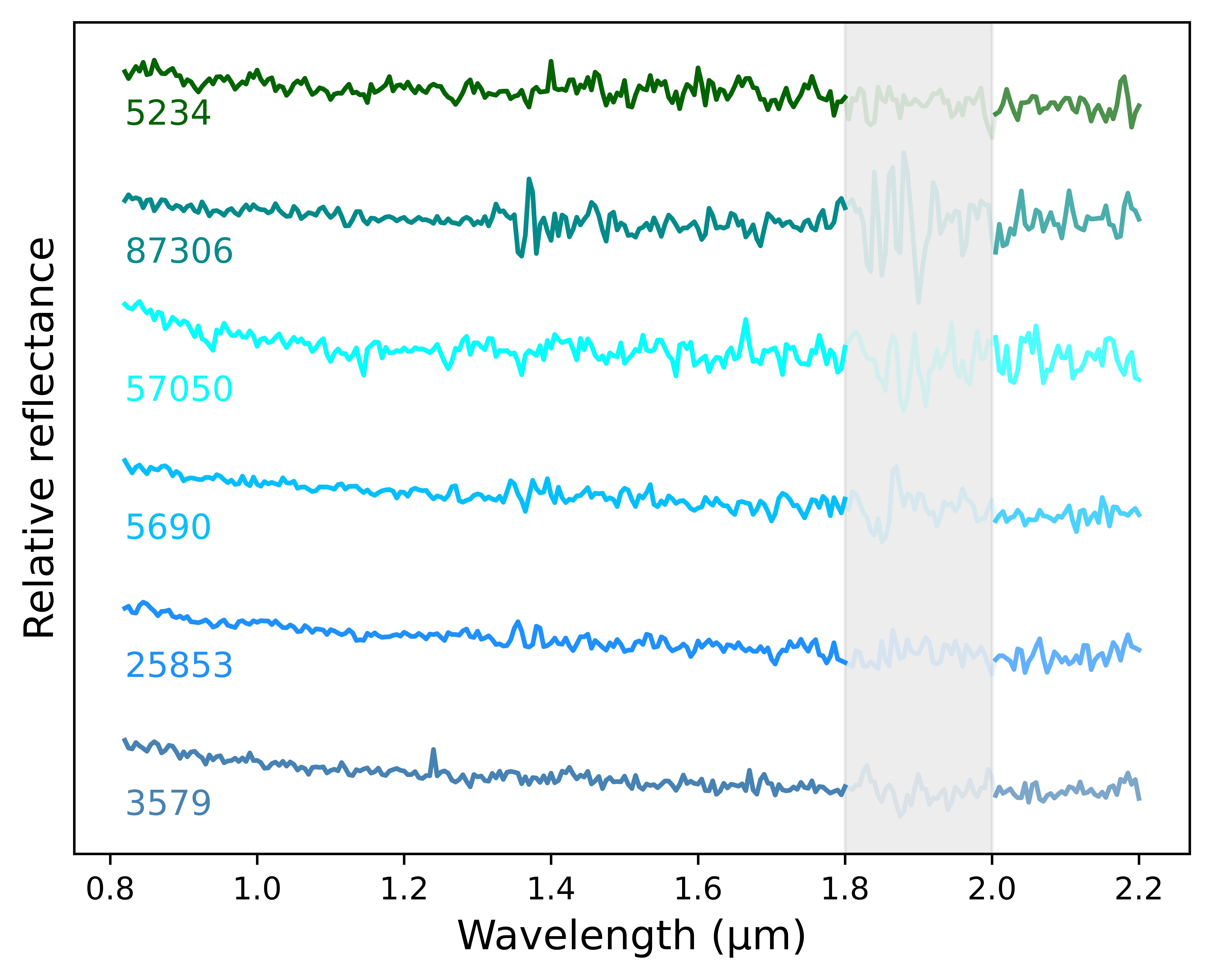}
    \includegraphics[width=8.95cm]{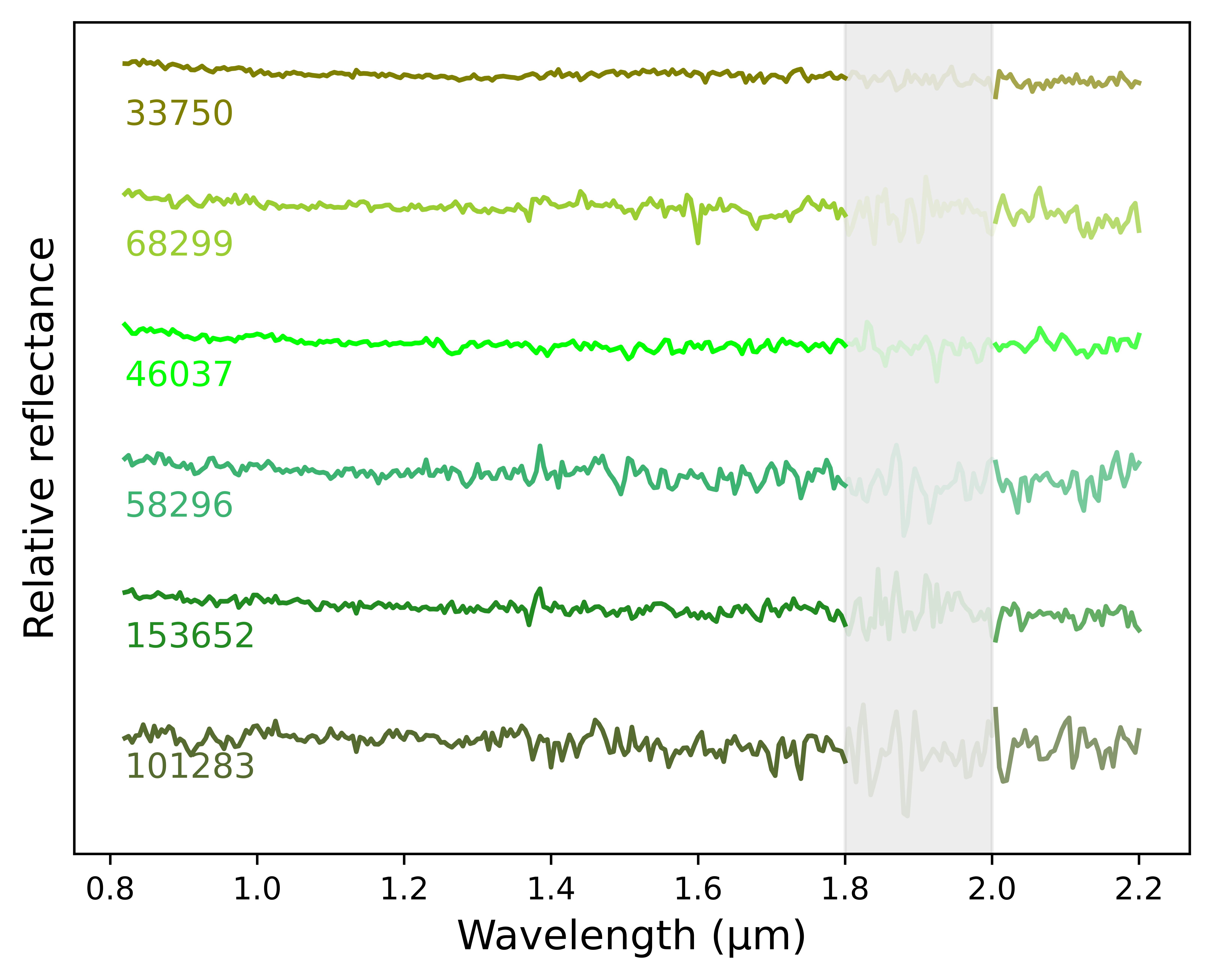}
    \includegraphics[width=8.95cm]{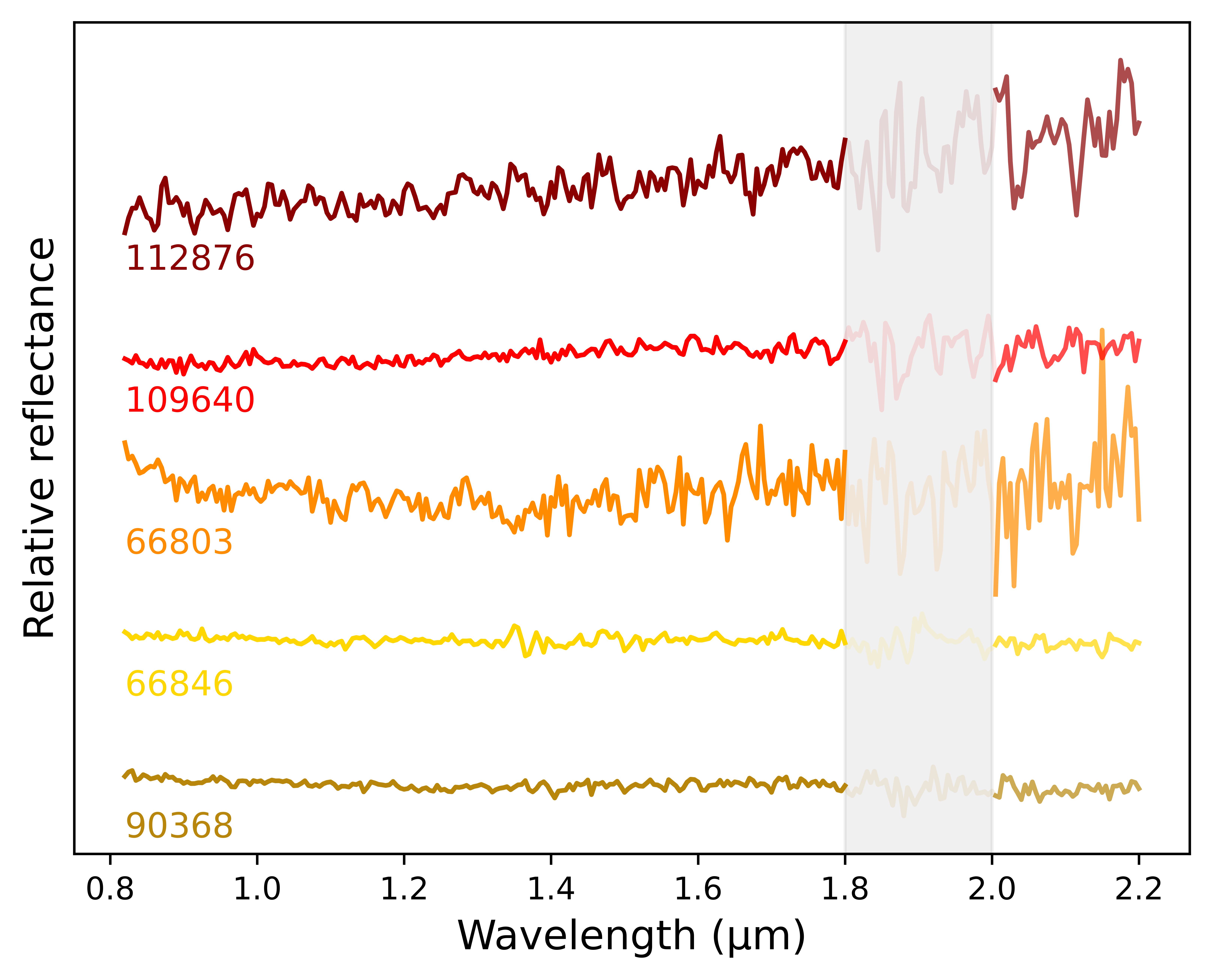}
    \caption{NIR spectral profiles of 23 asteroids normalized to unity at 1.0 $\mu$m. The spectra beyond 2.2 $\mu$m are truncated due to insufficient signal-to-noise. Note the spectra of objects (109640) and (112876), which exhibit comparatively redder slopes; as such, we identify them as possible outliers in the dataset. The gray zone marks the spectral region that is severely affected by atmospheric absorptions. The spectrum of Pallas is taken from \citep{SpectroscopyOfB-typeAsteroids}.
    }
    \label{fig:Spectra}
\end{figure*}

This study presents the largest NIR spectroscopic sample of the Pallas family to date, comprising 23 asteroids, including (2) Pallas. The NIR spectra of the Pallas family asteroids are shown in the wavelength range of 0.82--2.2 $\mu$m in Fig.~\ref{fig:Spectra} along with the spectrum of (2)~Pallas \citep[shown in black in the upper left panel]{SpectroscopyOfB-typeAsteroids}. We excluded spectral regions beyond 2.2 $\mu$m, where the spectra have low signal-to-noise ratios due to decreased spectrograph sensitivity and significant telluric absorption contamination. We computed the NIR spectral slope using linear regression with the linregress function from scipy.stats, which provides both the slope and its standard error. The slope was calculated in the wavelength range of 0.82--2.2 $\mu$m. The spectrum of (2) Pallas shows an overall negative spectral slope, with a steeper decline observed below 1.2 $\mu$m and a more gradual slope beyond that. No absorption features are observed in the spectrum of (2) Pallas. Most family members also show featureless bluish spectra, some exhibiting a change in the spectral slope between 1.0 and 1.2 $\mu$m. The Pallas family displays a tight spectral slope distribution, with a standard deviation of approximately 7 percent, suggesting a significant degree of homogeneity among its members. 
   
We classified the observed Pallas family asteroids using their NIR spectra, following the Bus-DeMeo taxonomy \citep{DeMeo:2009} which is based on principal component analysis of the spectral profile. Our results are listed in Table~\ref{tab:2}. Of the 23 observed targets, 19 are exclusively classified as B-types. Two asteroids with nearly neutral spectra could be classified as either B-type or Xn-type. (109640), with its slightly reddish spectrum, is classified as a Ch-type, while (112876) is a C-type asteroid. These two notable exceptions, (109640) and (112876), will be discussed in more detail in Section ~\ref{outliers}.

\subsection{Dynamical verification of outliers} \label{outliers}

\begin{figure}
    \centering
    \includegraphics[width=9cm]{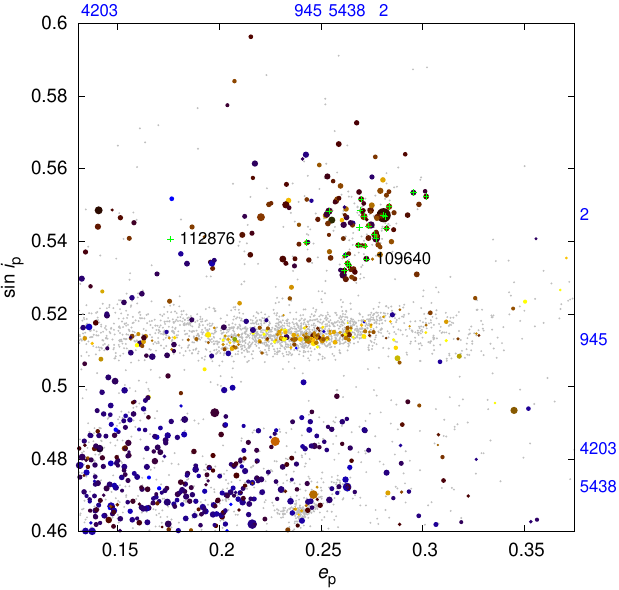}
    \caption{The proper eccentricity, $e_{\rm p}$, versus the proper inclination $\sin i_{\rm p}$ for the Pallas family and its broader surroundings. The locations of other families —(945) Barcelona, (4203) Brucato, and (5438) Lorre— are indicated on the right and top axes with blue numbers. Symbol sizes represent asteroid diameters. The colors (blue to yellow) indicate geometric albedos ranging from 0.05 to 0.25, obtained from WISE \citep{Wise2010}. Green crosses denote asteroids from this work, and gray symbols correspond to asteroids with no available size or albedo information. Two possible outliers, (109640) and (112876), are also indicated (black labels).}
    \label{ei_wise__lst}
\end{figure}

As shown in Fig.~\ref{fig:Spectra}, asteroids (109640) and (112876) exhibit distinct red spectra with positive spectral slopes. Given the tight spectral slope distribution among the family members, the two spectrally red objects could be interlopers. To further verify this possibility, we performed a dynamical analysis using their proper orbital elements following the methods by \cite{Broz:2008}. Fig.~\ref{ei_wise__lst} shows the positions of the possible outliers relative to the Pallas family and other asteroid families. (109640) is in the middle of $a$ (2.773 au) and $e$ but has a slightly lower $\sin i$, compared to the other members. Given that the background population density is low at high inclinations and the measured albedo of this object is $p_v$=0.150$\pm$0.016,  consistent with the family mean albedo of 0.141$\pm$ 0.043 \citep{2016AliLagoaPCFBtype}, we conclude that 109640 is not an outlier but a member of the Pallas family.

In contrast, (112876) is in the middle of $a$ (2.711 au) and $\sin i$, but at relatively low $e \simeq 0.17$, which is closer to that of (4203) Brucato ($e=0.15$). The Pallas family is located between the 8:3 and 5:2 mean-motion resonances with Jupiter at 2.72 au and 2.83 au, respectively \citep{ViolentHistory(2)Pallas}, whereas the Brucato family \citep{Carruba:2011} is located just below the 8:3 resonance. According to the dynamical simulations presented in \cite{ViolentHistory(2)Pallas}, this situation is suitable for transport by chaotic diffusion from Brucato to Pallas (and vice versa). This diffusive transport also explains the difference in inclination between (112876) and Brucato (Fig.~\ref{ei_wise__lst}). Therefore, we identify (112876) as an interloper, with the Brucato family being the most likely source of contamination.

\subsection{Physical properties of the Pallas family} \label{albedos}

Besides the NIR spectroscopy, we searched for other available observations of the Pallas family members at different wavelengths that probe various physical properties. The measured spectral slopes in the NIR, the available albedo and diameter measurements with WISE/NEOWISE \citep{Wise2010, Mainzer2014} as well as the optical spectral slopes based on the spectrophotometric observations with \textit{Gaia} \citep{GaiaCollaboration2023a} are listed in Table~\ref{tab:2}.

\begin{figure}
    \centering
    \includegraphics[width=9cm]{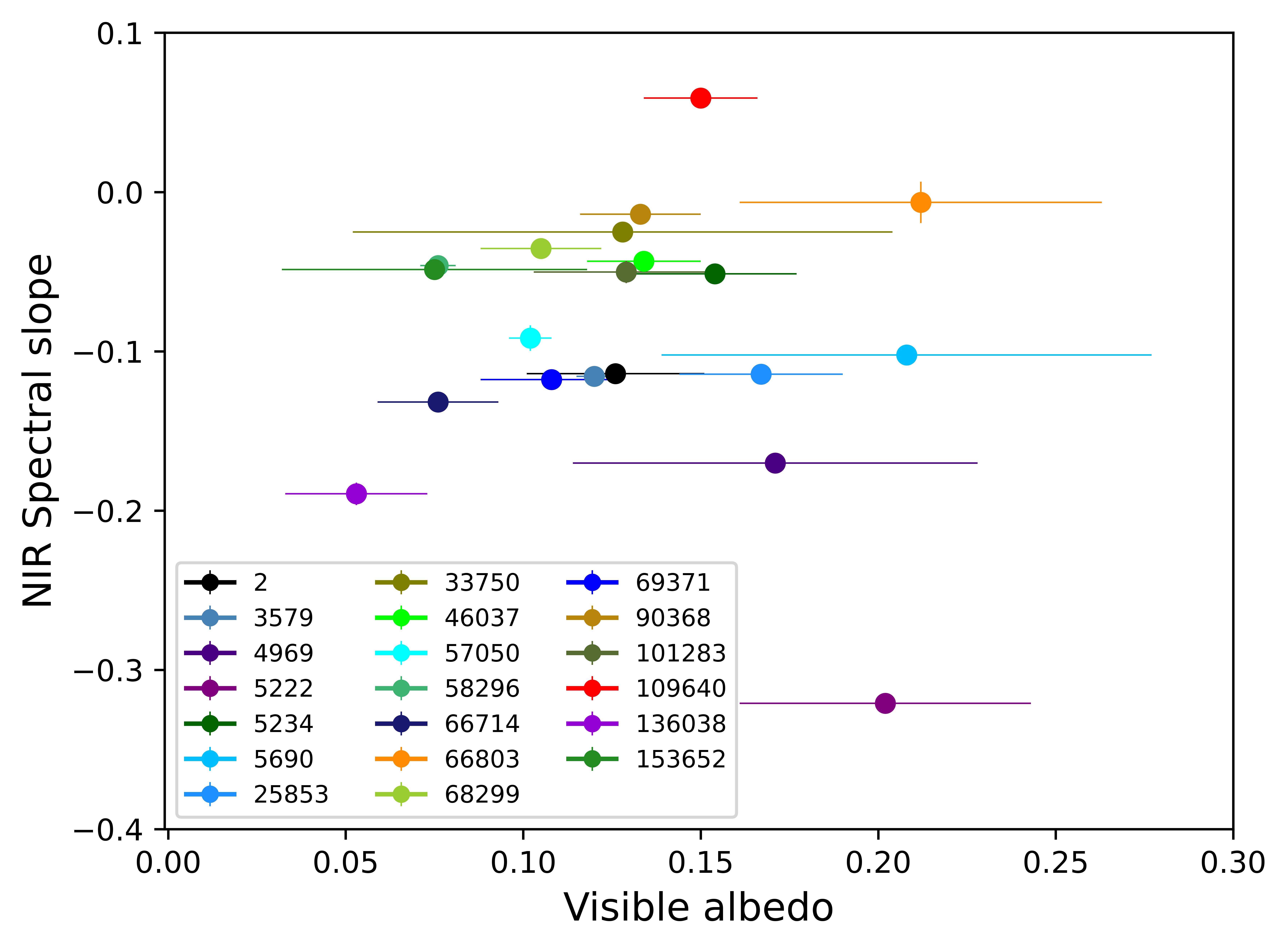}
    \caption{Visible albedo versus the NIR spectral slopes of observed Pallas family members. The albedo data is obtained from the 2019 \href{https://sbnarchive.psi.edu/pds4/non_mission/neowise_diameters_albedos_V2_0/data/}{NEOWISE diameters and albedos archive}.
    }
    \label{fig:VisalbedoVslope}
\end{figure}

Previous studies of main-belt asteroids, such as \cite{2014MasieroNIR}, suggest that high-albedo objects typically exhibit red slopes from visible to NIR wavelengths, while low-albedo objects tend to display blue slopes. This trend is particularly noticeable among high-albedo asteroids, such as S-types, which are mostly characterized by bright and red silicates. In contrast, asteroids with lower albedos, like those in the C-complex and D-types, show a broader variation in spectral slopes. The observed color-albedo correlation among main-belt asteroids is generally attributed to compositional differences across spectral types \citep{2014MasieroNIR}. As shown in Fig.~\ref{fig:VisalbedoVslope}, the visible albedo values of the observed Pallas family members show no apparent correlation with their NIR spectral slopes, likely due to the compositional homogeneity of the Pallas family.

\begin{figure}
    \centering
    \includegraphics[width=9cm]{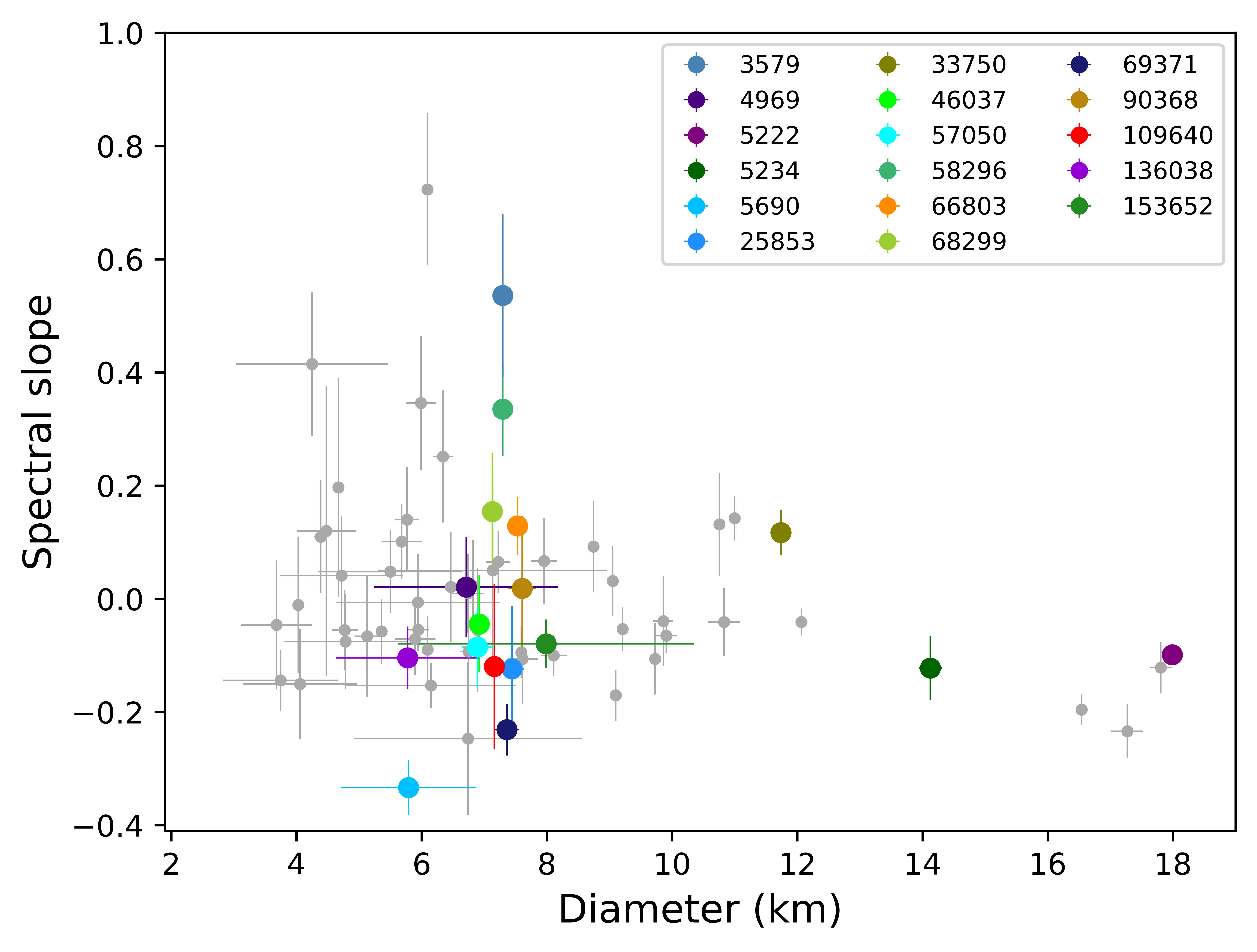}
    \caption{Spectral slope versus the size of the observed Pallas family members. The spectral slopes are calculated from \textit{Gaia}’s Data Release 3 (DR3) observations \citep{GaiaCollaboration:2016, GaiaCollaboration:2023b} and the sizes are obtained from the 2019 \href{https://sbnarchive.psi.edu/pds4/non_mission/neowise_diameters_albedos_V2_0/data/}{NEOWISE diameters and albedos archive}.}
    \label{fig:gaia}
\end{figure}

Based on analysis of the Sloan Digital Sky Survey (SDSS) data, \cite{Nesvorny2005} observed that asteroid families with the same taxonomical classification but at different ages demonstrate variations in color attributable to the space weathering effect. More recently, \cite{Thomas2021} analyzed the SDSS colors of C-complex families and identified two distinct trends, where the Hygiea-type trend shows a clear decrease in spectral slope (bluing) with increasing object size, reaching a minimum slope value, after which the slope increases (reddening) as object size continues to increase. The Themis-type trend has a clear increase in spectral slope (reddening) with increasing object size until a maximum slope value is reached. We first examined the NIR spectral slopes of 22 family members in relation to their derived diameters. The spectral slope was observed to decrease with increasing size for the three largest asteroids (D $>$ 10 km). However, among smaller asteroids (D $<$  10 km), a weaker but opposite correlation was observed. To verify the observed potential correlation, we expanded this study to include optical reflectance data for 77 Pallas family objects obtained from \href{https://cdn.gea.esac.esa.int/Gaia/gdr3/Solar_system/sso_reflectance_spectrum/}{\textit{Gaia} database}. We computed their optical spectral slope and associated uncertainties using linear regression over the wavelength range of 506–946 nm. The first and last few wavelength points were excluded due to low signal-to-noise ratios. Fig.~\ref{fig:gaia} shows the sizes versus the optical spectral slopes of the Pallas family asteroids. Asteroids observed in this study are highlighted in color, while other family members are shown in gray. Similar to patterns observed in other asteroid families, the smaller asteroids exhibit a much wider spectral slope distribution compared to their larger counterparts \citep{Parker:2008, Morate:2019, Tinaut-Ruano:2024}. However, we do not observe any significant correlation between the asteroid sizes and their spectral slopes.

\begin{center}
    \begin{table*}
    \caption{\label{tab:2}Physical and spectral properties of the Pallas family asteroids.}
    \centering
        \begin{tabular}{rcccccccccc}
        \hline\hline
        Asteroid & $H$ & $S'$ & $\sigma_{S'}$ & $D$ & $\sigma_{D}$ & $p_{V}$ & $\sigma_{p_{V}}$ & Taxonomy & \textit{Gaia} $S'$ & \textit{Gaia} $\sigma_{S'}$\\
         & (mag) & ($\mu{\rm m}^{-1}$) && (km) &&&&& ($\mu{\rm m}^{-1}$) &\\
        \hline
        \vrule height10pt width0pt
          2 &  4.12 & $-0.114$ & 0.004 & 513    & 3     & 0.126 & 0.025 & B    & $-0.074$ & 0.016\\
          3579 & 13.86 & $-0.116$ & 0.004 & 7.296  & 0.114 & 0.120 & 0.005 & B    & $+0.536$ & 0.145\\
          4969 & 12.92 & $-0.170$ & 0.005 & 6.712  & 1.472 & 0.171 & 0.057 & B    & $+0.021$ & 0.089\\
          5222 & 11.60 & $-0.095$ & 0.003 & 20.252 & 4.787 & 0.085 & 0.099 & B    & $-0.099$ & 0.016\\
          5234 & 12.28 & $-0.051$ & 0.006 & 14.125 & 0.183 & 0.154 & 0.023 & B    & $-0.122$ & 0.057\\
          5690 & 13.21 & $-0.102$ & 0.004 & 5.791  & 1.075 & 0.208 & 0.069 & B    & $-0.334$ & 0.048\\
         25853 & 13.57 & $-0.114$ & 0.004 & 7.443  & 0.189 & 0.167 & 0.023 & B    & $-0.124$ & 0.110\\
         33750 & 12.91 & $-0.025$ & 0.003 & 11.737 & 0.178 & 0.128 & 0.076 & B    & $+0.117$ & 0.039\\
         46037 & 14.01 & $-0.043$ & 0.003 & 6.918  & 0.142 & 0.134 & 0.016 & B    & $-0.044$ & 0.086\\
         57050 & 13.83 & $-0.092$ & 0.008 & 6.891  & 0.248 & 0.102 & 0.006 & B    & $-0.085$ & 0.071\\
         58296 & 14.26 & $-0.046$ & 0.006 & 7.295  & 0.141 & 0.076 & 0.005 & B    & $+0.335$ & 0.082\\
         66714 & 14.31 & $-0.132$ & 0.006 & 6.139  & 0.032 & 0.076 & 0.017 & B    & -        & -\\
         66803 & 13.00 & $-0.007$ & 0.013 & 7.533  & 0.143 & 0.212 & 0.051 & B    & $+0.129$ & 0.051\\
         66846 & 14.08 & $-0.008$ & 0.003 & -      & -     & -     & -     & B/Xn & $+0.089$ & 0.066\\
         68299 & 13.93 & $-0.035$ & 0.004 & 7.127  & 0.139 & 0.105 & 0.017 & B    & $+0.154$ & 0.103\\
         69371 & 13.77 & $-0.118$ & 0.006 & 7.363  & 0.193 & 0.108 & 0.020 & B    & $-0.231$ & 0.046\\
         87306 & 13.44 & $-0.061$ & 0.007 & -      & -     & -     & -     & B    & $-0.153$ & 0.059\\
         90368 & 13.82 & $-0.014$ & 0.003 & 7.605  & 0.22  & 0.133 & 0.017 & B/Xn & $+0.018$ & 0.128\\
        101283 & 14.23 & $-0.050$ & 0.007 & 6.137  & 0.122 & 0.129 & 0.026 & B    & -        & -\\
        109640 & 14.00 & $+0.059$ & 0.004 & 7.157 & 0.087 & 0.150  & 0.016 & Ch & -0.120 & 0.145\\
        136038 & 15.01 & $-0.189$ & 0.007 & 5.776  & 1.141 & 0.053 & 0.02  & B    & $-0.104$ & 0.055\\
        153652 & 13.77 & $-0.049$ & 0.004 & 7.987  & 2.357 & 0.075 & 0.043 & B    & $-0.079$ & 0.043\\
        \hline
        \end{tabular}
        \tablefoot{$H$ is the absolute magnitude, $S'$ is the spectral slope, $D$ is the diameter, $p_{V}$ is the visible albedo whereas $\sigma_{S'}$, $\sigma_{D}$ and $\sigma_{p_{V}}$ are the uncertainties in spectral slope, diameter, and visible albedo, respectively. The data is obtained from the AKARI IRC asteroid flux catalog by \cite{AKARI2018alilagoa}, and from the recent 2019 NEOWISE diameters and albedos archive, taken by the WISE Camera instrument of Wide-Field Infrared Survey Explorer \citep{Wise2010}.}
    \end{table*}
\end{center}

\subsection{Meteorite analogs for the Pallas family}

\begin{figure*}
    \centering
    \includegraphics[width=8.95cm]{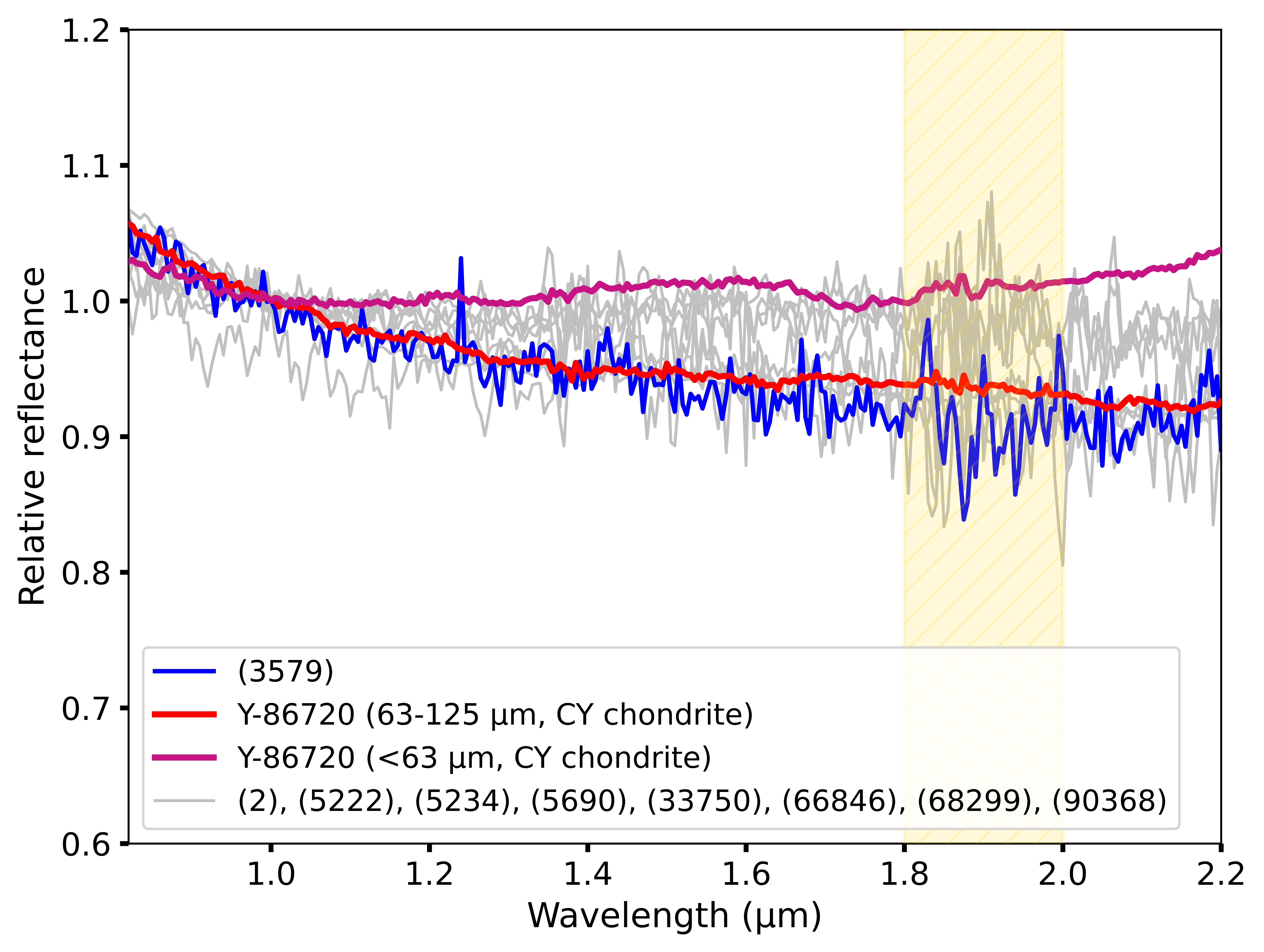}
    \includegraphics[width=8.95cm]{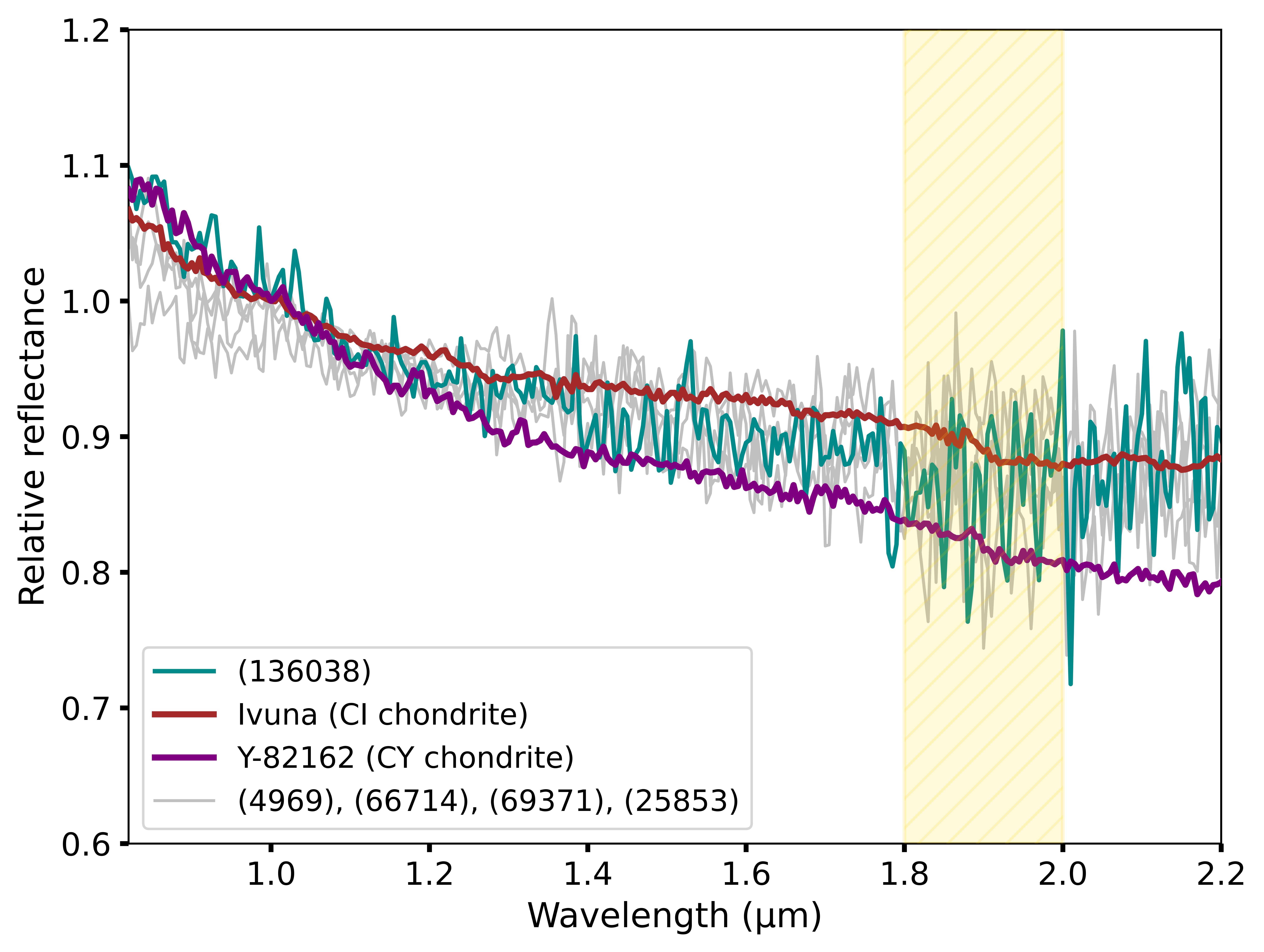}
    \includegraphics[width=8.95cm]{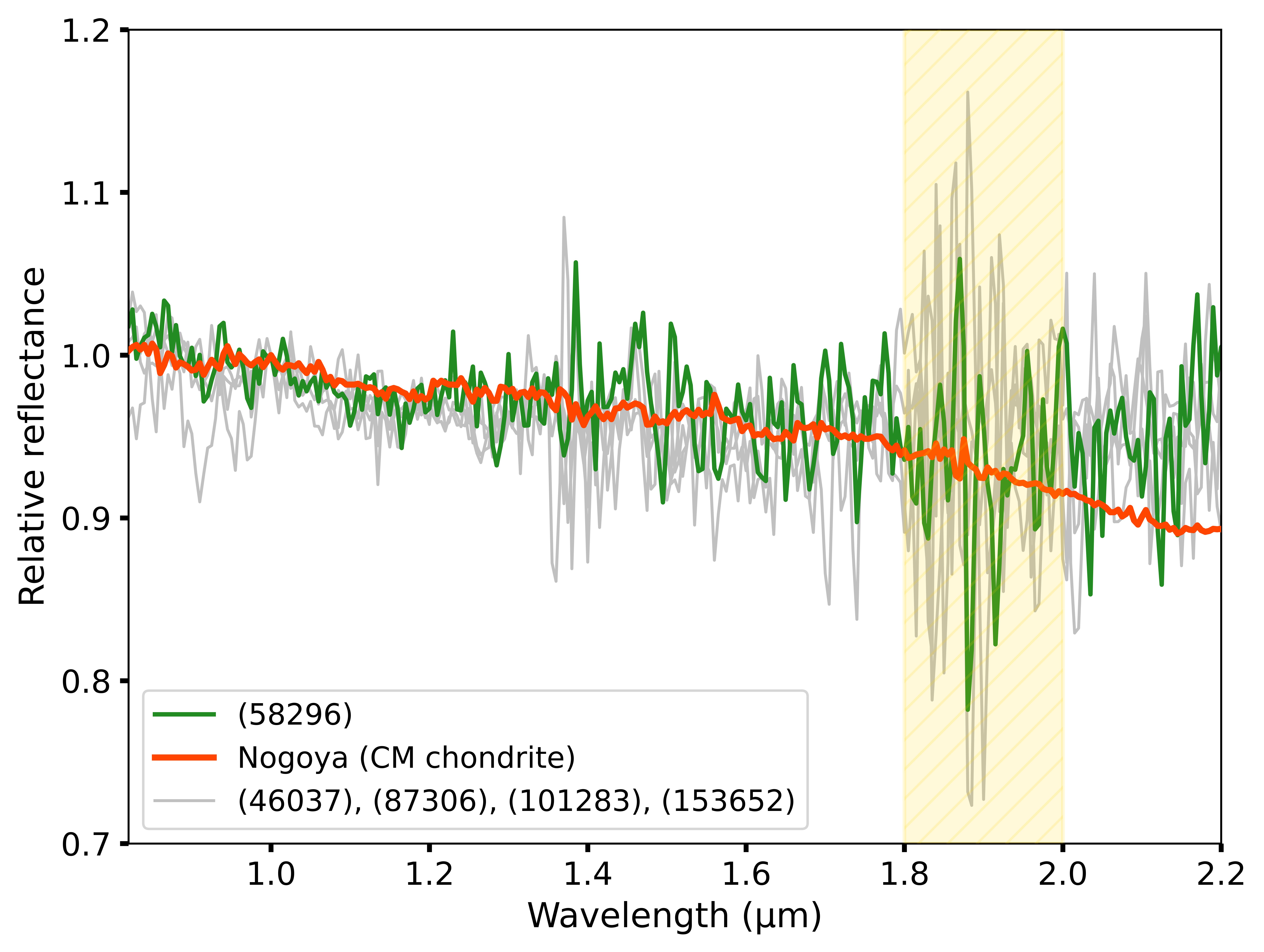}
    \includegraphics[width=8.95cm]{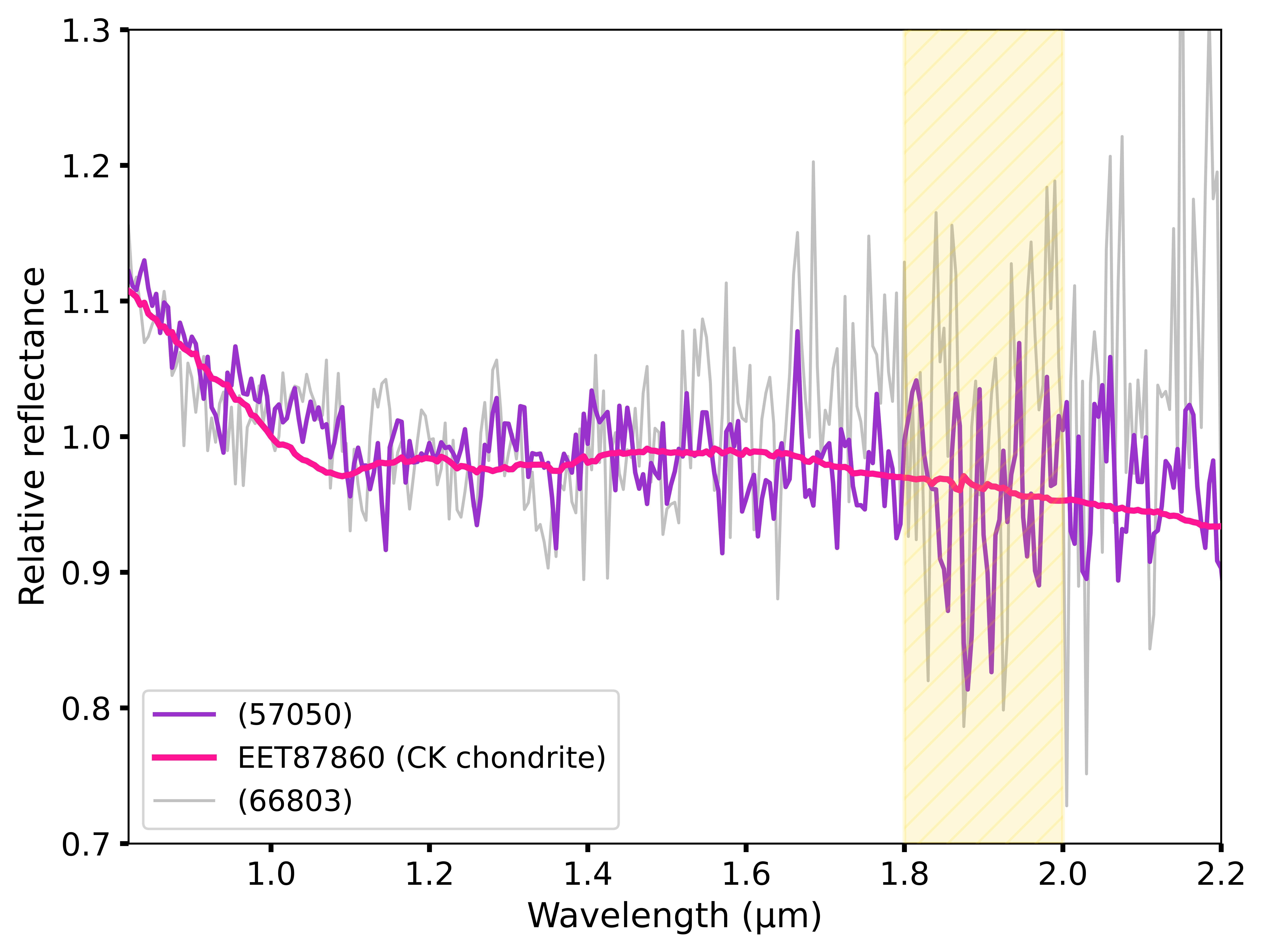}
    \caption{Four groups of asteroids with their respective best-fit meteorite analogs. The meteorite spectra were obtained from the RELAB library \citep{RELAB, Milliken:2020}, and the asteroid spectra are from this work. A representative asteroid from each group is shown in color, while the remaining asteroids in the same meteorite-matching group are displayed in gray. In the legend of the top-left plot, the values 63–125 $\mu$m and <63 $\mu$m indicate the grain sizes of the corresponding meteorite spectra. The yellow zone marks the spectral region that is severely affected by atmospheric absorptions.
    }
    \label{fig:AstMeteo}
\end{figure*}

To establish connections between asteroid spectra and meteorite analogs and to further study the possible compositions and surface properties of the Pallas family asteroids, we used the RELAB online database hosted by Brown University \citep{RELAB}. We searched for analogs among carbonaceous meteorite reflectance spectra and set the wavelength range the same as the asteroid spectra (0.82--2.2 $\mu$m). We used natural meteorite samples that had not been artificially heated or irradiated in the lab. The samples of these spectra consist of rocky particulates with sizes smaller than 125 $\mu$m, with measurements conducted at a phase angle of 30$^{\circ}$. We performed a chi-square test to measure the goodness of fit between the meteorite reflectance and the asteroid reflectance spectra. This method has been adopted in previous studies to establish asteroid-meteorite connections (e.g., \citealt{SpectroscopyOfB-typeAsteroids, ConnectingAstMeteo2022DeMeo}). We categorized the asteroid spectra according to the meteorite group they matched. Fig.~\ref{fig:AstMeteo} shows the following four categories:
\begin{itemize}
\item (i) Asteroid 3579 represents a group of asteroids that can be best matched to the spectra of the Y-86720 CY chondrite with different particle sizes. Specifically, (2) Pallas, 5222, 5234, and 5690 resemble Y-86720 CY chondrite (c1mb20c, grain size: 63-125 $\mu$m), while 33750, 66846, 68299, and 90368 match Y-86720 CY chondrite (c1mb20b, grain size: <63 $\mu$m).
\item(ii) Asteroid 136038, representing asteroids 4969, 66714, 69371, and 25853, shows the closest spectral match to the Ivuna CI chondrite (c4mb60, grain size: <125 $\mu$m) as well as to the Y-82162 CY chondrite (cbmb19, chip).
\item(iii) Asteroid 58296, along with asteroids 46037, 87306, 101283, and 153652, shows the closest resemblance to the Nogoya CM chondrite (c4mb62, grain size: 63-125 $\mu$m). 
\item(iv) Asteroids 57050 and 66803 show the best match with EET87860 CK chondrite (c1mp04, grain size: <63 $\mu$m). Their spectra exhibit a decreasing reflectance with increasing wavelengths up to 1.1 $\mu$m, after which they remain relatively flat or neutral.
\end{itemize}

\subsection{Comparison between Phaethon and the Pallas family members}

\begin{figure}
    \centering
    \includegraphics[width=8.95cm]{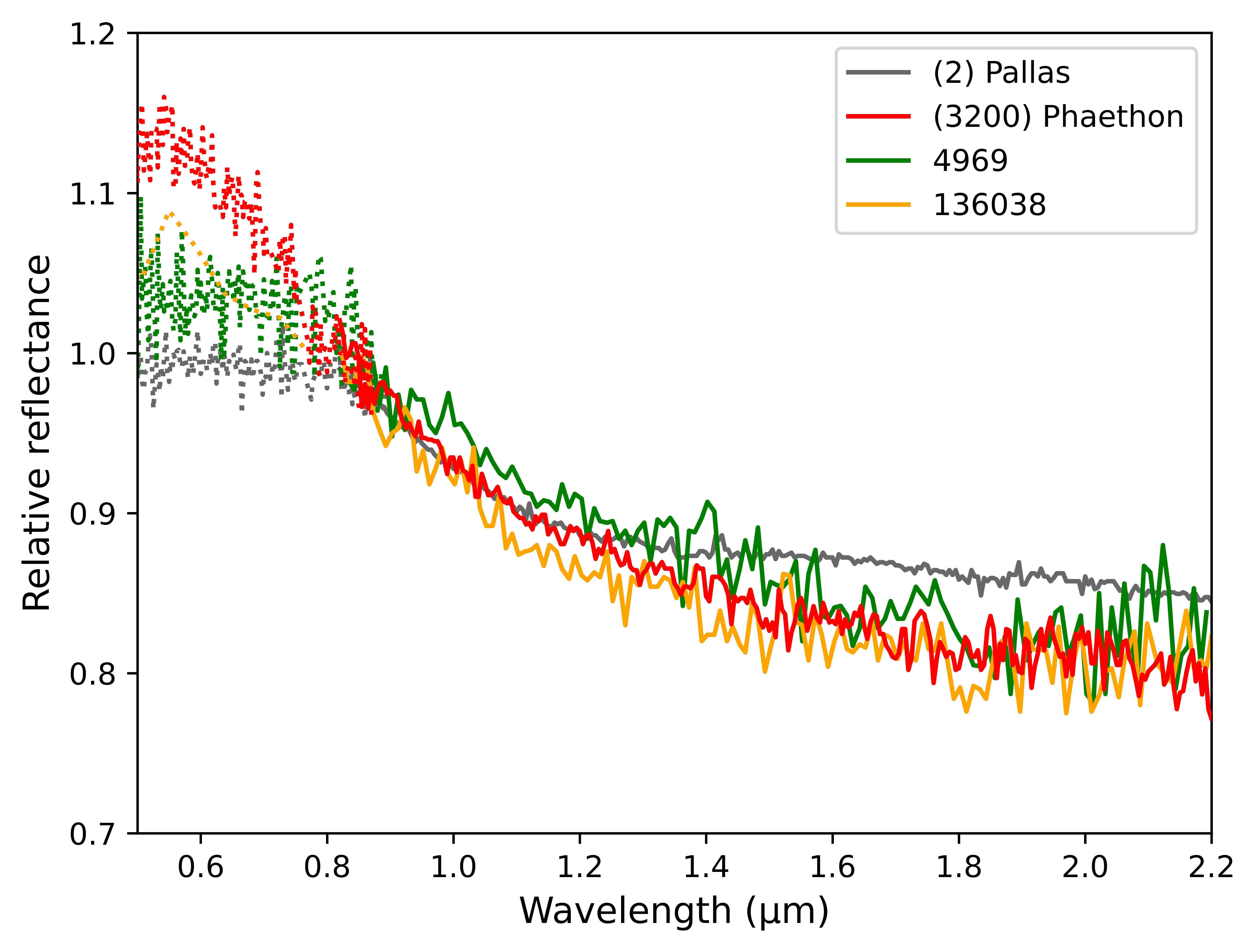}
    \caption{Spectral comparison between asteroid Phaethon, Pallas, and similar Pallas family members, (4969) and (136038). The Phaethon VIS-NIR reflectance spectrum (red) is obtained from \href{http://smass.mit.edu/data/smass/smassneoref8/}{Small Main-belt Asteroid Spectroscopic Survey (SMASS)}\citep{Binzel:2004} and \href{http://smass.mit.edu/data/spex/sp34/}{MIT-UH-IRTF Survey}\citep{Binzel:2006}, the visible spectra of Pallas (gray dots) and 4969 (green dots) are obtained from \href{http://smass.mit.edu/data/smass/smass2/}{SMASS II}\citep{Bus1999}. The Pallas NIR spectrum (gray line) is obtained from \cite{SpectroscopyOfB-typeAsteroids}, the 136038 VIS spectrum (yellow dots) is obtained from \textit{Gaia} DR3 database \citep{GaiaCollaboration:2016, GaiaCollaboration:2023b} and the rest are from this work. All spectra were normalized to unity at 0.82$\,\mu$m.}
    \label{fig:phaethon}
\end{figure}

As previously noted, it has been widely acknowledged that the spectrum of (3200) Phaethon shares notable similarities with that of (2) Pallas in the optical and near-infrared \citep{OriginofPhaethon}. However, Pallas ($513\pm 3$\,km) is orders of magnitude larger than Phaethon ($5.1\pm0.2$ km; \citealt{Phaethonsize}), and Phaethon exhibits a noticeably bluer spectral slope than Pallas in the optical \citep{OriginofPhaethon}. As shown in Fig.~\ref{fig:Spectra}, some small Pallas family asteroids exhibit significantly bluer spectra compared to (2)~Pallas. To further investigate the potential link between Phaethon and the Pallas family, we compared its spectrum with those in our sample over a wide range of wavelengths from 0.4 to 2.2\,$\mu$m. We took the visible spectrum of Phaethon from Small Main-belt Asteroid Spectroscopic Survey (SMASS) \citep{Binzel:2004} and its NIR spectrum from MIT-UH-IRTF survey \citep{Binzel:2006}. The visible spectra of Pallas and (4969) were obtained from SMASS II \citep{Bus1999}. Additionally, the visible spectrum of (136038) was retrieved from the \textit{Gaia} DR3 database \citep{GaiaCollaboration:2016, GaiaCollaboration:2023b}. We found that two Pallas family asteroids, (4969) ($6.7\pm1.5$\,km) and (136038) ($5.8\pm1.1$\,km), show a nearly identical spectral match with Phaethon in the NIR, as shown in Fig.~\ref{fig:phaethon}. However, we note that these asteroids exhibit significantly different spectral slopes from Phaethon at shorter wavelengths below 0.8\,$\mu$m. These differences highlight the importance of extending the analysis of spectral properties across a wider wavelength range to gain a comprehensive understanding of the complex resurfacing processes on Phaethon.

\section{Discussion} \label{sec:discussion}

Unlike other sizable B-type asteroid families, such as the Themis family, which exhibit a wide range of spectral variability \citep{Florczak:1999, Fornasier:2016}, our NIR observations reveal that the Pallas family has a more tightly confined spectral slope distribution, with more than 80\% of its observed members classified as B-type. In addition, \cite{2016AliLagoaPCFBtype} reported that members of the Pallas family consistently show higher albedos than non-family B-types. The highly similar NIR spectra and albedo measurements of the Pallas family underscore its compositional homogeneity, in line with previous studies, although based on a much smaller sample \citep{SpectroscopyOfB-typeAsteroids}.

Previous studies have identified that B-type asteroids show a highly diverse spectral behavior in the NIR \citep{SpectroscopyOfB-typeAsteroids, BinMagnetite, NIRbtypeast}. Uniquely, all the Pallas B-type asteroids exhibit blue spectral slopes in the NIR. \cite{BeckPoch2021} investigated color variations as a function of asteroid size for C- and B-type primitive asteroids based on the SDSS database. They observed that smaller objects ($D<20\,{\rm km}$) appear bluer than larger ones. The authors discuss a potential cause for this observed trend: smaller asteroids are hypothesized to have rock-dominated or fragmented surfaces, while larger asteroids have surfaces composed of finer, smaller grains. However, as shown in Table~\ref{tab:2} and Fig.~\ref{fig:gaia}, we observed a greater dispersion among smaller family members, and our observations, along with those from \textit{Gaia}, do not support the aforementioned hypothesis.  Specifically, 50\% of the sampled smaller objects exhibit redder spectral characteristics compared to larger asteroids such as (5222) (see Fig.~\ref{fig:gaia}). This suggests that object size is not a significant factor contributing to the blueness of the spectral slopes in Pallas family asteroids.

To understand the origin of the puzzling blueness, \cite{SpectroscopyOfB-typeAsteroids} investigated potential connections between the spectra of B-type asteroids and meteorite analogs. They found that Pallas-like asteroids (7~objects) were spectrally similar to the CK, CV, or CO meteorite groups. In this study, using a significantly larger sample (22~objects), we observe that the spectra of most Pallas family members are consistent with those of CY and CI meteorites, while a small fraction resembles CM and CK meteorites.

CY chondrites are rare carbonaceous meteorites recovered from Antarctica, with mineralogical, textural, and elemental characteristics that fall between those of CI and CM chondrites \citep{CY-King}. They show evidence of aqueous alteration on their parent bodies and contain dehydrated phyllosilicates. Furthermore, CY meteorites contain highly abundant Fe-sulfides that are sulfur-depleted and predominantly in the form of troilite (FeS) \citep{MacLennan_2024NatAs...8...60M}. Laboratory studies by \cite{possibleexplanation2022Bslope} suggest that troilite contributes significantly to reducing spectral slopes, leading to the "bluing" effect. 

However, troilite is not the only agent responsible for the blue color of carbonaceous meteorites. In the case of CI chondrites, particularly the Ivuna meteorite, the blue spectral slope has been linked to the presence of magnetite and graphite \citep{BinMagnetite, Clark2011}. \cite{PropertiesOfCMCloutis} suggest that both magnetite and organic content in CM chondrites account, at least in part, for the overall blue slope. The blue slope observed in many CK spectra is likely attributable to the presence of Fe-bearing spinels and/or magnetite \citep{PropertiesOfCKCloutis}.

The comparative study between the Pallas family asteroids and meteorite analogs consistently indicates that the parent body of the Pallas family experienced aqueous alteration. However, different meteorite analogs lead to different levels of metamorphism of the parent body. For instance, CI chondrites are considered among the most primitive materials in the Solar System and have never experienced significant heating since their formation \citep{Weisberg2006}. In contrast, studies indicate that CY meteorites experienced peak temperatures exceeding 500\,$^\circ{\rm C}$, possibly as high as 700–800\,$^\circ{\rm C}$ \citep{CY-King}, leading to the thermal decomposition of phyllosilicates such as serpentine and saponite, and the formation of troilite from pyrrhotite.

We propose the following scenario to explain the varying levels of metamorphism deduced from the meteorite analog analysis among family members. The parent body of the Pallas family initially consisted of CI-like materials and experienced a moderate amount of aqueous alteration and partial differentiation, which left the surface enriched with magnetite and salts \citep{ViolentHistory(2)Pallas}. During the cratering event, some of the ejecta were heated to peak temperatures well above 500\,$^\circ{\rm C}$ (or 800\,K). Consistent with this, \cite{Wakita-impact} conducted three-dimensional oblique impact simulations and demonstrated that impact-induced shock heating during collisions could raise temperatures above 873\,K, sufficient to cause the decomposition of phyllosilicates. Evidence of shock heating has been observed in various meteorite samples \citep{Nakamura:2005, Nakato:2008, Abreu:2013}. In this scenario, objects displaying CY-like spectra are those that experienced significant impact heating. Ejecta that experienced moderate heating are more similar to CM and CK chondrites. Finally, fragments that escaped the heating process retained their CI-like composition.

Additionally, the surfaces of airless planetary bodies are exposed to micrometeoroid impacts and bombardment from the solar wind, and their optical properties can be modified by the space weathering effect \citep{Hapke:2001}. Space weathering has been relatively well studied for S-type asteroids, which generally leads to spectral darkening and reddening due to the presence and concentration of nanophase iron particles \citep{Clark:2002, Brunetto_2006Icar..184..327B}. However, the modification of surface properties due to radiation and ion bombardment is not well understood for primitive C-complex asteroids. Experimental studies on ion irradiation of carbonaceous chondrites have produced inconsistent results, including bluing and brightening \citep{Moroz:2004, Lantz:2017}, reddening and darkening \citep{Keller:2015}, or no significant changes \citep{Lantz:2015}. \cite{Lantz_2018Icar..302...10L} reported that the spectral modifications of primitive asteroids depend on the albedo and composition of the surface material. \cite{Trang:2021} found that hydrated iron-bearing phyllosilicates must be present on the surface to enable the formation of nanophase magnetite, which causes a carbonaceous asteroid to appear spectrally blue.

Lastly, grain size is identified as another factor that can influence spectral slopes. Studies, such as \cite{Cantillo-2023}, suggest that an increase in grain size leads to a decrease in the spectral slope, resulting in a bluer appearance. While larger grains can reduce overall reflectance, this trend is not observed in our meteorite samples. This indicates that the observed spectral characteristics are more likely influenced by compositional differences rather than variations in grain size.


\subsection{Potential link between Phaethon and Pallas family}

The near-Earth asteroid (3200) Phaethon holds significant scientific importance as the first asteroid identified as the parent body of a meteor shower. It exhibits comet-like activity as well as a distinct blue spectral slope \citep{Jewitt:2006, Licandro:2007}. Phaethon’s orbit brings it extremely close to the Sun, making it a natural laboratory for studying thermal processes and space weathering effects \citep{Hanus:2016, MacLennan:2021}. Therefore, it has been extensively studied, with previous research attempting to establish a link between Phaethon and main-belt asteroid families through analyses of spectral similarities and dynamical pathways. For instance, \cite{OriginofPhaethon} proposed a spectral and dynamical connection between Phaethon and the Pallas family, though with a low delivery efficiency of approximately 2\% from the main belt to a Phaethon-like orbit. \cite{Todorovic:2018} investigated the influence of the two strongest mean motion resonances (MMRs) with Jupiter, namely the 8:3 and 5:2 MMRs, in the region of the Pallas family and revisited the potential dynamical linkage between Phaethon and Pallas through numerical simulations. They found a highly efficient dynamical flow between Pallas and Phaethon, with nearly 50\% of test particles being transported from the MMRs to Phaethon-like orbits. However, other studies suggest that the inner or central main asteroid belt is a more likely source region for NEAs such as Phaethon \citep{Bottke:2002}. Building on previous work, \cite{MacLennan:2021} proposed a link between Phaethon’s source and inner belt families, such as the (329) Svea \citep{Morate:2019} or (142) Polana family \citep{Walsh:2013, Pinilla-Alonso:2016}.
More recently, \cite{Broz:2024} demonstrated that the Polana family is the most important source delivering CI-like materials to near-Earth orbits. They used a set of 56 asteroid families, together with their spectroscopic classifications, their observed size-frequency distributions (SFDs), and also suitable collisional models. While the Polana family is currently producing most NEOs ($\sim$57.7 bodies), their probabilistic model also predicts up to 3.5 bodies larger than 1 km originating from the Pallas family, suggesting that (2) Pallas remains a potential source for objects like Phaethon. \cite{Knezevic:2024} analyzed the potential dynamical origins of Phaethon, considering the Pallas family in the outer belt and the (329) Svea and (142) Polana families in the inner belt. Contrary to earlier findings, the authors concluded that the Pallas family is more likely to be the source region for Phaethon. Their analysis demonstrated that the Pallas family has at least twice the probability of delivering asteroids to Phaethon’s orbital vicinity compared to the two inner belt families. The discrepancy among dynamical analyses may be due to differences in initial condition assumptions and the irreversibility of chaos, as backward integration of chaotic orbits may not provide reliable solutions for understanding their dynamic history.

\begin{figure}
    \centering
    \includegraphics[width=8.95cm]{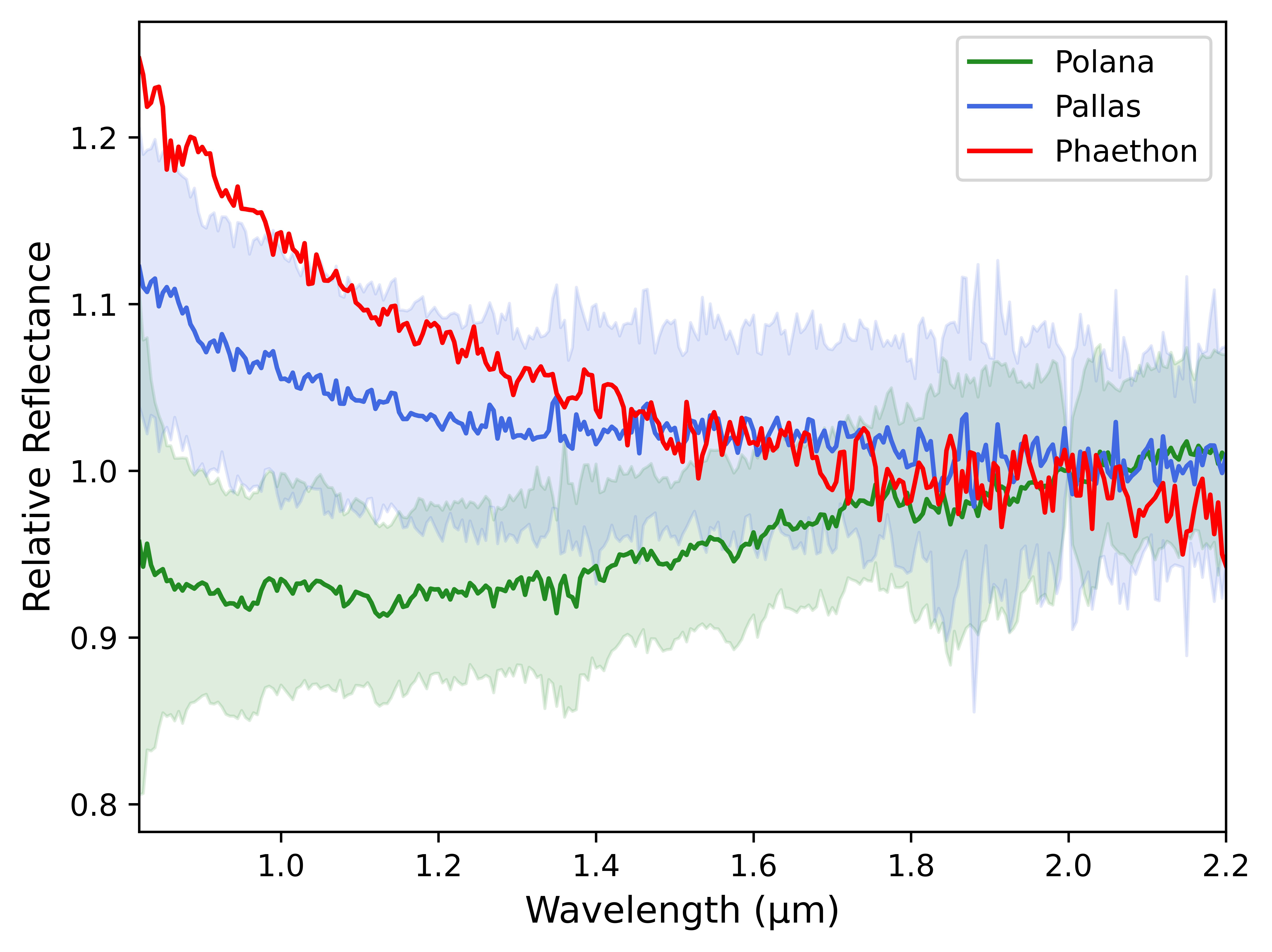}
    \caption{Spectral comparison between asteroid Phaethon, Pallas, and Polana family. The Phaethon NIR reflectance spectrum (red) is obtained from \href{http://smass.mit.edu/data/spex/sp34/}{MITHNEOS MIT-Hawaii Near-Earth Object Spectroscopic Survey}\citep{OriginofPhaethon}, the mean spectrum of Pallas family (blue) is calculated from this work and the mean spectrum of Polana family (green) is obtained from \href{https://ui.adsabs.harvard.edu/link_gateway/2024pds..data..114P/doi:10.26033/YMJ5-4084}{PRIMASS-L Spectra Bundle V2.0}\citep{Pinilla-Alonso:2016}. All spectra were normalized to unity at 2.0$\,\mu$m.}
    \label{fig:polana}
\end{figure}
In our analysis, Fig.~\ref{fig:phaethon} shows that two small Pallas family asteroids, (4969) and (136038), which are comparable in size to Phaethon, exhibit nearly identical spectral profiles to Phaethon in the NIR. Approximately 10\% of the sampled Pallas family members exhibit this remarkable similarity, reinforcing the previously noted connection between Phaethon and the Pallas family.
Furthermore, we compared the mean spectra and 1$\sigma$ standard deviations of the Pallas and Polana families with Phaethon to determine which family exhibits a higher degree of spectral similarity to the asteroid. As illustrated in Fig.~\ref{fig:polana}, Phaethon exhibits a significantly closer spectral resemblance to the Pallas family than to the Polana family. The calculated slope of Phaethon is $-0.155\pm0.003$ $\mu{\rm m}^{-1}$, while the slope of the mean spectrum of the Pallas family and Polana family is $-0.056\pm0.002$ $\mu{\rm m}^{-1}$ and $0.07\pm0.001$ $\mu{\rm m}^{-1}$, respectively. More importantly, the Polana family exhibits a broad concave feature centered around 1.0\,$\mu$m, which has been attributed to Fe-bearing phyllosilicates \citep{PropertiesOfCOCloutis, BinMagnetite}. This absorption feature is clearly absent in the spectra of both Phaethon and the Pallas family. Given the resilience of such spectral features, it is unlikely that they could be entirely erased by surface modification processes, such as space weathering \citep{Rivkin:2015, MacLennan_2024NatAs...8...60M}. Our observations support the hypothesis that Pallas is more likely to be the parent body of Phaethon compared to Polana.

The remarkable spectral similarity between small Pallas family asteroids and Phaethon offers valuable clues to better understand not only the origin of this NEO but also the mechanisms behind its distinct blue color. Previously, the pronounced blueness of Phaethon’s spectrum has been proposed to be due to its close proximity to the Sun, where temperatures nearing 1100\,K drive mass-loss activity and the thermal sublimation of surface organic materials \citep{Lisse2022}. However, our study shows that very similar blue spectra are also observed among kilometer-sized Pallas family asteroids that have never been within 1.75\,au of the Sun, with their surface temperatures never exceeding 300\,K. Our observations suggest that high temperatures and thermal alteration are not necessarily the primary drivers of Phaethon's distinct blue spectral slope, emphasizing the importance of further investigation to determine the mechanisms behind its spectral blueness.

Most asteroid studies are currently limited to the optical and NIR regions, where few absorption features exist, and compositional information is largely inferred from spectral slope analysis. However, as discussed earlier, spectral slopes can be influenced by factors such as space weathering and the grain size of surface particles \citep{Cantillo-2023, Lopez-Oquendo:2024}, introducing uncertainties in these interpretations. While observing in the UV, the 3-$\mu$m region, and the mid-IR is challenging; these regions contain key diagnostic signatures that are less susceptible to surface processes than spectral slopes \citep{Tatsumi:2022, Tatsumi:2023, Tinaut-Ruano:2024}. Expanding spectral studies to these regions is therefore crucial for obtaining more accurate and robust information about the composition and mineralogy of asteroids. Phaethon has been reported to lack the 3-$\mu$m absorption band that is commonly observed among hydrated asteroids, including Pallas \citep{Takir:2020}. The absence of the sharp absorption feature can be due to the thermal evolution of Phaethon’s surface or the limited sensitivity of ground-based observations. Future observations of Phaethon, along with (4969) and (136038), using advanced facilities such as the \textit{James Webb Space Telescope} (JWST) \citep{Wright:2023}, have the potential to enable a more detailed and comprehensive comparison between Phaethon and the Pallas family. Moreover, the upcoming JAXA mission, DESTINY+ \citep{Arai:2024}, is expected to provide high-quality data on Phaethon’s surface composition and spectral properties. The combined efforts of JWST observations and measurements from DESTINY+ are anticipated to yield definitive evidence regarding Phaethon’s origin.

\section{Summary}

We conducted a detailed analysis of NIR spectroscopy for the largest sample to date of 23 Pallas family asteroids. This study investigates potential correlations between spectral slopes and other physical parameters, such as size and albedo. Additionally, we explore the spectral connections between these asteroids and their meteorite analogs, revisiting the hypothesized link between the Pallas family and the near-Earth asteroid (3200) Phaethon. The analysis yielded the following key findings:

\begin{enumerate}
    \item The NIR spectra of the Pallas family asteroids exhibit consistent spectral profiles, suggesting a compositional homogeneity inherited from the parent body. Minor variations in spectral slopes likely reflect differences in the extent of alteration experienced by individual members.

    \item Most family members closely match CY and CI meteorites, with a subset showing affinities to CM and CK meteorites.
    
    \item The blue spectral slopes observed are likely due to the presence of magnetite, troilite, graphite, phyllosilicates, and ferrihydrite—minerals formed through aqueous alteration.

    \item The remarkable spectral similarity between Phaethon and certain Pallas family members of similar sizes supports the hypothesis that Phaethon originated within the Pallas family.
    
    \item Approximately 10\% of kilometer-sized Pallas family members show similar spectral characteristics that resemble those of Phaethon. Therefore, the distinct blueness in Phaethon’s spectrum does not appear to be attributed solely to high surface temperatures or thermal alteration.

\end{enumerate}

\begin{acknowledgements}
The authors were Visiting Astronomers at the Infrared Telescope Facility, which is operated by the University of Hawaii under contract 80HQTR24DA010 with the National Aeronautics and Space Administration (NASA). This work has used spectra acquired by investigators at the NASA RELAB facility at Brown University.
The Czech Science Foundation has supported this work through grants 25-16507S (M.~Bro\v z) and 22-17783S (J.~Hanu\v s). We would like to thank Bill Bottke for his constructive discussions. We would also like to thank Noemi Pinilla-Alonso for providing the spectra of the Polana family. Finally, we would like to thank the referee, Julia de León, for the thoughtful comments and constructive suggestions.
\end{acknowledgements}

\bibliography{paper}
\bibliographystyle{aa}

\end{document}